\documentclass[zpreprint,zbstpl]{zeus_paper}
\usepackage[english]{babel}
\newcommand{\Zdetdesc}{%
A detailed description of the ZEUS detector can be found 
elsewhere~\cite{zeus:1993:bluebook}. A brief outline of the 
components that are most relevant for this analysis is given
below.\xspace}

\newcommand{\Zsttdesc}[1]{%
The STT consisted of 48 sectors of two different sizes. Each sector
contained 192 (small sector) or 264 (large sector) straws of diameter
7.5 mm arranged into 3 layers. The sectors were trapezoidal in shape
and each subtended an azimuthal angle of $60^{\circ}$ -- 6 sectors
formed a so-called superlayer. A particle passing through the complete
detector traversed 8 superlayers, which were rotated around the beam
direction at angles of $30^{\circ}$ or $15^{\circ}$ to each other. The STT
covered the polar-angle region $5^{\circ}<\theta<23^{\circ}$.
}
\newcommand{\Zcaldesc}{%
The high-resolution uranium--scintillator calorimeter (CAL)~\citeCAL consisted 
of three parts: the forward (FCAL), the barrel (BCAL) and the rear (RCAL)
calorimeters. Each part was subdivided transversely into towers and
longitudinally into one electromagnetic section (EMC) and either one (in RCAL)
or two (in BCAL and FCAL) hadronic sections (HAC). The smallest subdivision of
the calorimeter is called a cell.  
The CAL energy resolutions, as measured under
test-beam conditions, are $\sigma(E)/E=0.18/\sqrt{E}$ for electrons and
$\sigma(E)/E=0.35/\sqrt{E}$ for hadrons, with $E$ in $\Gev$.}

\newcommand{\Zbacdesc}{%
The iron yoke surrounding the CAL was instrumented with proportional
drift chambers to form the Backing Calorimeter (BAC)~\citeBAC.  The
BAC consisted of 5142 aluminium chambers inserted into the gaps
between $7.3\,{\rm cm}$ thick iron plates (10, 9 and 7 layers in 
forward, central (barrel) and rear subdetectors, respectively) 
serving as calorimeter absorber.
The chambers were typically $5\,{\rm m}$ long and had a wire spacing of
$1.5\,{\rm cm}$. The anode wires were covered by $50\,{\rm cm}$ long cathode
pads. The BAC was equipped with energy
readout and position-sensitive readout for muon tracking.  The former
was based on 1692 pad towers ($50 \times 50\,{\rm cm^2}$), providing an
energy resolution $\sigma(E)/E = 1.0/\sqrt E$, where $E$ is expressed
in \gev. The position information
from the wires allowed the reconstruction of muon trajectories in two
dimensions ($XY$ in barrel and $YZ$ in endcaps) with a spatial accuracy
of a few mm.}
\chardef\usc=95
\chardef\til=126
\catcode`\@=11 
\DeclareRobustCommand\xdotspace{\futurelet\@let@token\@xdotspace}
\def\@xdotspace{%
  \ifx\@let@token.\else
  \ifx\@let@token\bgroup.\else
  \ifx\@let@token\egroup.\else
  \ifx\@let@token\/.\else
  \ifx\@let@token\ .\else
  \ifx\@let@token~.\else
  \ifx\@let@token!.\else
  \ifx\@let@token,.\else
  \ifx\@let@token:.\else
  \ifx\@let@token;.\else
  \ifx\@let@token?.\else
  \ifx\@let@token/.\else
  \ifx\@let@token'.\else
  \ifx\@let@token).\else
  \ifx\@let@token-.\else
  \ifx\@let@token\@xobeysp.\else
  \ifx\@let@token\space.\else
  \ifx\@let@token\@sptoken.\else
   .\space
   \fi\fi\fi\fi\fi\fi\fi\fi\fi\fi\fi\fi\fi\fi\fi\fi\fi\fi}
\catcode`\@=12 

\newcommand{\stru}[2]{%
   \relax\ifmmode\hbox{\vrule height#1 depth#2 width0pt}%
   \else\vrule height#1 depth#2 width0pt\fi}

\newcommand{\Ronum}[1]{\uppercase\expandafter{\romannumeral#1}}
\newcommand{\ronum}[1]{\expandafter{\romannumeral#1}}
\DeclareRobustCommand{\LaTeXZ}{%
  \LaTeX\kern-.05em4\kern-.1em
  {\raisebox{-0.2ex}{$\scriptstyle\text{ZEUS}$}}\xspace}

\DeclareMathAlphabet{\mathbf}{OT1}{cmr}{bx}{sl}
\newcommand{\eVdist}{\kern-0.06667em}

\newcommand{\Gev}{{\text{Ge}\eVdist\text{V\/}}}

\newcommand{\gev}{{\,\text{Ge}\eVdist\text{V\/}}}

\newcommand{\Tesla}{\,\text{T}}

\newcommand{\slashfrac}[2]{%
  \raisebox{0.5ex}{\ensuremath #1}\kern-0.12em/\kern-0.08em
  \raisebox{-.8ex}{\ensuremath #2}}

\newcommand{\sqr}[3]{%
    {\vcenter{\hrule height.#3ex\hbox{\vrule width.#2ex height#1ex
     \kern#1ex\vrule width.#3ex}\hrule height.#2ex}}}

\catcode`\@=11 
\newcommand{\parenbar}{\mathpalette\p@renb@r}
\def\p@renb@r#1#2{\vbox{%
  \ifx#1\scriptscriptstyle \dimen@.7em\dimen@ii.2em\else
  \ifx#1\scriptstyle \dimen@.8em\dimen@ii.25em\else
  \dimen@1em\dimen@ii.4em\fi\fi \offinterlineskip
  \ialign{\hfill##\hfill\cr
    \vbox{\hrule width\dimen@ii}\cr
    \noalign{\vskip-.3ex}%
    \hbox to\dimen@{$\mathchar300\hfil\mathchar301$}\cr
    \noalign{\vskip-.3ex}%
    $#1#2$\cr}}}
\catcode`\@=12 

\newcommand{\IP}{{\rm I$\kern-0.01667em$P}\xspace}

\mathchardef\qsm=63
\mathchardef\pls=43
\mathchardef\mns=512
\mathchardef\plm=518
\mathchardef\eql=61
\mathchardef\smallleft=300
\mathchardef\smallright=301
\mathchardef\les=316
\mathchardef\gre=318
\mathchardef\leq=532
\mathchardef\grq=533

\catcode`\@=11 
\newcounter{pict@width}
\newcounter{pict@height}
\newlength{\pict@scale}
\newcommand{\psfigadd}[4]{%
\setcounter{pict@width}{1*\ratio{#2+\pict@scale/2}{\pict@scale}}
\setcounter{pict@height}{1*\ratio{#3+\pict@scale/2}{\pict@scale}}
\setlength{\unitlength}{\pict@scale}
\hbox to #2{\hspace{-\fill}\begin{picture}(\thepict@width,\thepict@height)
\put(0,0){\psfig{figure=#1,width=#2,height=#3,clip=}}
\SetScale{0.283466457}
\SetWidth{1.763889}
{#4}
\end{picture}}
}
\newcounter{pict@widthfst}
\newcounter{pict@widthscd}
\newcounter{pict@widthtot}
\newcommand{\psfigaddtwo}[7]{%
\setcounter{pict@widthfst}{1*\ratio{#2+\pict@scale/2}{\pict@scale}}
\setcounter{pict@widthscd}{1*\ratio{#2+#4+\pict@scale/2}{\pict@scale}}
\setcounter{pict@widthtot}{1*\ratio{#2+#4+#6+\pict@scale/2}{\pict@scale}}
\setcounter{pict@height}{1*\ratio{#3+\pict@scale/2}{\pict@scale}}
\setlength{\unitlength}{\pict@scale}
\hbox{\hspace{-\fill}\begin{picture}(\thepict@widthtot,\thepict@height)
\put(0,0){\psfig{figure=#1,width=#2,height=#3,clip=}}
\put(\thepict@widthscd,0){\psfig{figure=#5,width=#6,height=#3,clip=}}
\SetScale{0.283466457}
\SetWidth{1.763889}
{#7}
\end{picture}}
}
\newcommand{\psfigror}[4]{%
\setcounter{pict@width}{1*\ratio{#2+\pict@scale/2}{\pict@scale}}
\setcounter{pict@height}{1*\ratio{#3+\pict@scale/2}{\pict@scale}}
\setlength{\unitlength}{\pict@scale}
\hbox{\begin{picture}(\thepict@width,\thepict@height)
\put(0,\thepict@height){\psfig{figure=#1,width=#3,height=#2,clip=,angle=270}}
\SetScale{0.283466457}
\SetWidth{1.763889}
{#4}
\end{picture}}
}
\newcommand{\psfigrol}[4]{%
\setcounter{pict@width}{1*\ratio{#2+\pict@scale/2}{\pict@scale}}
\setcounter{pict@height}{1*\ratio{#3+\pict@scale/2}{\pict@scale}}
\setlength{\unitlength}{\pict@scale}
\hbox{\begin{picture}(\thepict@width,\thepict@height)
\put(0,0){\psfig{figure=#1,width=#3,height=#2,clip=,angle=90}}
\SetScale{0.283466457}
\SetWidth{1.763889}
{#4}
\end{picture}}
}
\catcode`\@=12 
\newlength\listtextwidth

\catcode`\@=11 
\newlength{\@tabfninsert}
\newlength{\@tabfnwidth}
\newcommand{\tabfootnote}[2]{%
  \setlength{\@tabfninsert}{0.8em}
  \setlength{\@tabfnwidth}{\textwidth}
  \addtolength{\@tabfnwidth}{-\@tabfninsert}
  \addtolength{\@tabfnwidth}{-0.4em}
  \noindent\makebox[\@tabfninsert][r]{\footnotesize$^{#1}$\hfil}\hfill%
  \parbox[t]{\@tabfnwidth}{\footnotesize #2\hfill}}
\catcode`\@=12 

\def\citeRubinskyMalka{{\cite{%
thesis:rubinsky:2009,*thesis:malka:2011%
}}\xspace}
\def\citeCTD{{\cite{%
nim:a279:290,*npps:b32:181,*nim:a338:254%
}}\xspace}
\def\citeMVD{{\cite{%
nim:a581:656%
}}\xspace}

\def\citeCAL{{\cite{%
nim:a309:77,*nim:a309:101,*nim:a321:356,*nim:a336:23%
}}\xspace}

\def\cite6mT{{\cite{%
thesis:gosau:2007%
}}\xspace}
\def\citeBAC{{\cite{%
nim:a300:480%
}}\xspace}

\begin{document}
\title{
Measurement of the {\boldmath $t$} dependence in exclusive photoproduction
of {\boldmath $\Upsilon(1S)$} mesons at HERA
}

\author{ZEUS Collaboration}
\draftversion{3.2 (postreading)}
\prepnum{DESY--11--186}
\date{24 October 2011}

\abstract{
The exclusive photoproduction reaction $\gamma\,p \rightarrow
\Upsilon(1S) \,p$ has been studied with the ZEUS detector in $ep$ collisions
at HERA using an integrated luminosity of 468~pb$^{-1}$. 
The measurement covers the kinematic range $60<W<220$~\gev~and $Q^2<1$
\gev$^2$, where $W$ is the photon--proton centre-of-mass energy and
$Q^2$ is the photon virtuality. 
The exponential slope, $b$, of the $t$ dependence of the cross section,
where $t$ is the squared four-momentum transfer at the proton vertex,
has been measured, yielding 
$b=4.3^{+2.0}_{-1.3}$ (stat.)$\, ^{+0.5}_{-0.6}$ (syst.)~\gev$^{-2}$.
This constitutes the first measurement of the $t$ dependence of 
the $\gamma\,p \rightarrow \Upsilon(1S) \,p$ cross section.
}

\makezeustitle

\clearpage
%
%
%
%



\def\3{\ss}
\pagenumbering{Roman}

                                                   %
\begin{center}
  {\Large  The ZEUS Collaboration}
\end{center}

{\small


        {\raggedright
H.~Abramowicz$^{45, ah}$, 
I.~Abt$^{35}$, 
L.~Adamczyk$^{13}$, 
M.~Adamus$^{54}$, 
R.~Aggarwal$^{7, c}$, 
S.~Antonelli$^{4}$, 
P.~Antonioli$^{3}$, 
A.~Antonov$^{33}$, 
M.~Arneodo$^{50}$, 
V.~Aushev$^{26, 27, z}$
% ...
        }

\newpage


\makebox[3em]{$^{1}$}
\begin{minipage}[t]{14cm}
{\it Argonne National Laboratory, Argonne, Illinois 60439-4815, USA}~$^{A}$

\end{minipage}\\
\makebox[3em]{$^{2}$}
\begin{minipage}[t]{14cm}
{\it Andrews University, Berrien Springs, Michigan 49104-0380, USA}

\end{minipage}\\
\makebox[3em]{$^{3}$}
\begin{minipage}[t]{14cm}
{\it INFN Bologna, Bologna, Italy}~$^{B}$

\end{minipage}\\
\makebox[3em]{$^{4}$}
\begin{minipage}[t]{14cm}
{\it University and INFN Bologna, Bologna, Italy}~$^{B}$

\end{minipage}\\
\makebox[3em]{$^{5}$}
\begin{minipage}[t]{14cm}
{\it Physikalisches Institut der Universit\"at Bonn,
Bonn, Germany}~$^{C}$

\end{minipage}\\
\makebox[3em]{$^{6}$}
\begin{minipage}[t]{14cm}
{\it H.H.~Wills Physics Laboratory, University of Bristol,
Bristol, United Kingdom}~$^{D}$

\end{minipage}\\
\makebox[3em]{$^{7}$}
\begin{minipage}[t]{14cm}
{\it Panjab University, Department of Physics, Chandigarh, India}

\end{minipage}\\
\makebox[3em]{$^{8}$}
\begin{minipage}[t]{14cm}
{\it Calabria University,
Physics Department and INFN, Cosenza, Italy}~$^{B}$

\end{minipage}\\
\makebox[3em]{$^{9}$}
\begin{minipage}[t]{14cm}
{\it Institute for Universe and Elementary Particles, Chonnam National University,\\
Kwangju, South Korea}

\end{minipage}\\
\makebox[3em]{$^{10}$}
\begin{minipage}[t]{14cm}
{\it Jabatan Fizik, Universiti Malaya, 50603 Kuala Lumpur, Malaysia}~$^{E}$

\end{minipage}\\
\makebox[3em]{$^{11}$}
\begin{minipage}[t]{14cm}
{\it Nevis Laboratories, Columbia University, Irvington on Hudson,
New York 10027, USA}~$^{F}$

\end{minipage}\\
\makebox[3em]{$^{12}$}
\begin{minipage}[t]{14cm}
{\it The Henryk Niewodniczanski Institute of Nuclear Physics, Polish Academy of \\
Sciences, Krakow, Poland}~$^{G}$

\end{minipage}\\
\makebox[3em]{$^{13}$}
\begin{minipage}[t]{14cm}
{\it AGH-University of Science and Technology, Faculty of Physics and Applied Computer
Science, Krakow, Poland}~$^{H}$

\end{minipage}\\
\makebox[3em]{$^{14}$}
\begin{minipage}[t]{14cm}
{\it Department of Physics, Jagellonian University, Cracow, Poland}

\end{minipage}\\
\makebox[3em]{$^{15}$}
\begin{minipage}[t]{14cm}
{\it Deutsches Elektronen-Synchrotron DESY, Hamburg, Germany}

\end{minipage}\\
\makebox[3em]{$^{16}$}
\begin{minipage}[t]{14cm}
{\it Deutsches Elektronen-Synchrotron DESY, Zeuthen, Germany}

\end{minipage}\\
\makebox[3em]{$^{17}$}
\begin{minipage}[t]{14cm}
{\it INFN Florence, Florence, Italy}~$^{B}$

\end{minipage}\\
\makebox[3em]{$^{18}$}
\begin{minipage}[t]{14cm}
{\it University and INFN Florence, Florence, Italy}~$^{B}$

\end{minipage}\\
\makebox[3em]{$^{19}$}
\begin{minipage}[t]{14cm}
{\it Fakult\"at f\"ur Physik der Universit\"at Freiburg i.Br.,
Freiburg i.Br., Germany}

\end{minipage}\\
\makebox[3em]{$^{20}$}
\begin{minipage}[t]{14cm}
{\it School of Physics and Astronomy, University of Glasgow,
Glasgow, United Kingdom}~$^{D}$

\end{minipage}\\
\makebox[3em]{$^{21}$}
\begin{minipage}[t]{14cm}
{\it Department of Engineering in Management and Finance, Univ. of
the Aegean, Chios, Greece}

\end{minipage}\\
\makebox[3em]{$^{22}$}
\begin{minipage}[t]{14cm}
{\it Hamburg University, Institute of Experimental Physics, Hamburg,
Germany}~$^{I}$

\end{minipage}\\
\makebox[3em]{$^{23}$}
\begin{minipage}[t]{14cm}
{\it Imperial College London, High Energy Nuclear Physics Group,
London, United Kingdom}~$^{D}$

\end{minipage}\\
\makebox[3em]{$^{24}$}
\begin{minipage}[t]{14cm}
{\it Institute of Particle and Nuclear Studies, KEK,
Tsukuba, Japan}~$^{J}$

\end{minipage}\\
\makebox[3em]{$^{25}$}
\begin{minipage}[t]{14cm}
{\it Institute of Physics and Technology of Ministry of Education and
Science of Kazakhstan, Almaty, Kazakhstan}

\end{minipage}\\
\makebox[3em]{$^{26}$}
\begin{minipage}[t]{14cm}
{\it Institute for Nuclear Research, National Academy of Sciences, Kyiv, Ukraine}

\end{minipage}\\
\makebox[3em]{$^{27}$}
\begin{minipage}[t]{14cm}
{\it Department of Nuclear Physics, National Taras Shevchenko University of Kyiv, Kyiv, Ukraine}

\end{minipage}\\
\makebox[3em]{$^{28}$}
\begin{minipage}[t]{14cm}
{\it Kyungpook National University, Center for High Energy Physics, Daegu,
South Korea}~$^{K}$

\end{minipage}\\
\makebox[3em]{$^{29}$}
\begin{minipage}[t]{14cm}
{\it Institut de Physique Nucl\'{e}aire, Universit\'{e} Catholique de Louvain, Louvain-la-Neuve,\\
Belgium}~$^{L}$

\end{minipage}\\
\makebox[3em]{$^{30}$}
\begin{minipage}[t]{14cm}
{\it Departamento de F\'{\i}sica Te\'orica, Universidad Aut\'onoma
de Madrid, Madrid, Spain}~$^{M}$

\end{minipage}\\
\makebox[3em]{$^{31}$}
\begin{minipage}[t]{14cm}
{\it Department of Physics, McGill University,
Montr\'eal, Qu\'ebec, Canada H3A 2T8}~$^{N}$

\end{minipage}\\
\makebox[3em]{$^{32}$}
\begin{minipage}[t]{14cm}
{\it Meiji Gakuin University, Faculty of General Education,
Yokohama, Japan}~$^{J}$

\end{minipage}\\
\makebox[3em]{$^{33}$}
\begin{minipage}[t]{14cm}
{\it Moscow Engineering Physics Institute, Moscow, Russia}~$^{O}$

\end{minipage}\\
\makebox[3em]{$^{34}$}
\begin{minipage}[t]{14cm}
{\it Moscow State University, Institute of Nuclear Physics,
Moscow, Russia}~$^{P}$

\end{minipage}\\
\makebox[3em]{$^{35}$}
\begin{minipage}[t]{14cm}
{\it Max-Planck-Institut f\"ur Physik, M\"unchen, Germany}

\end{minipage}\\
\makebox[3em]{$^{36}$}
\begin{minipage}[t]{14cm}
{\it NIKHEF and University of Amsterdam, Amsterdam, Netherlands}~$^{Q}$

\end{minipage}\\
\makebox[3em]{$^{37}$}
\begin{minipage}[t]{14cm}
{\it Physics Department, Ohio State University,
Columbus, Ohio 43210, USA}~$^{A}$

\end{minipage}\\
\makebox[3em]{$^{38}$}
\begin{minipage}[t]{14cm}
{\it Department of Physics, University of Oxford,
Oxford, United Kingdom}~$^{D}$

\end{minipage}\\
\makebox[3em]{$^{39}$}
\begin{minipage}[t]{14cm}
{\it INFN Padova, Padova, Italy}~$^{B}$

\end{minipage}\\
\makebox[3em]{$^{40}$}
\begin{minipage}[t]{14cm}
{\it Dipartimento di Fisica dell' Universit\`a and INFN,
Padova, Italy}~$^{B}$

\end{minipage}\\
\makebox[3em]{$^{41}$}
\begin{minipage}[t]{14cm}
{\it Department of Physics, Pennsylvania State University, University Park,\\
Pennsylvania 16802, USA}~$^{F}$

\end{minipage}\\
\makebox[3em]{$^{42}$}
\begin{minipage}[t]{14cm}
{\it Polytechnic University, Sagamihara, Japan}~$^{J}$

\end{minipage}\\
\makebox[3em]{$^{43}$}
\begin{minipage}[t]{14cm}
{\it Dipartimento di Fisica, Universit\`a 'La Sapienza' and INFN,
Rome, Italy}~$^{B}$

\end{minipage}\\
\makebox[3em]{$^{44}$}
\begin{minipage}[t]{14cm}
{\it Rutherford Appleton Laboratory, Chilton, Didcot, Oxon,
United Kingdom}~$^{D}$

\end{minipage}\\
\makebox[3em]{$^{45}$}
\begin{minipage}[t]{14cm}
{\it Raymond and Beverly Sackler Faculty of Exact Sciences, School of Physics, \\
Tel Aviv University, Tel Aviv, Israel}~$^{R}$

\end{minipage}\\
\makebox[3em]{$^{46}$}
\begin{minipage}[t]{14cm}
{\it Department of Physics, Tokyo Institute of Technology,
Tokyo, Japan}~$^{J}$

\end{minipage}\\
\makebox[3em]{$^{47}$}
\begin{minipage}[t]{14cm}
{\it Department of Physics, University of Tokyo,
Tokyo, Japan}~$^{J}$

\end{minipage}\\
\makebox[3em]{$^{48}$}
\begin{minipage}[t]{14cm}
{\it Tokyo Metropolitan University, Department of Physics,
Tokyo, Japan}~$^{J}$

\end{minipage}\\
\makebox[3em]{$^{49}$}
\begin{minipage}[t]{14cm}
{\it Universit\`a di Torino and INFN, Torino, Italy}~$^{B}$

\end{minipage}\\
\makebox[3em]{$^{50}$}
\begin{minipage}[t]{14cm}
{\it Universit\`a del Piemonte Orientale, Novara, and INFN, Torino,
Italy}~$^{B}$

\end{minipage}\\
\makebox[3em]{$^{51}$}
\begin{minipage}[t]{14cm}
{\it Department of Physics, University of Toronto, Toronto, Ontario,
Canada M5S 1A7}~$^{N}$

\end{minipage}\\
\makebox[3em]{$^{52}$}
\begin{minipage}[t]{14cm}
{\it Physics and Astronomy Department, University College London,
London, United Kingdom}~$^{D}$

\end{minipage}\\
\makebox[3em]{$^{53}$}
\begin{minipage}[t]{14cm}
{\it Faculty of Physics, University of Warsaw, Warsaw, Poland}

\end{minipage}\\
\makebox[3em]{$^{54}$}
\begin{minipage}[t]{14cm}
{\it National Centre for Nuclear Research, Warsaw, Poland}

\end{minipage}\\
\makebox[3em]{$^{55}$}
\begin{minipage}[t]{14cm}
{\it Department of Particle Physics and Astrophysics, Weizmann
Institute, Rehovot, Israel}

\end{minipage}\\
\makebox[3em]{$^{56}$}
\begin{minipage}[t]{14cm}
{\it Department of Physics, University of Wisconsin, Madison,
Wisconsin 53706, USA}~$^{A}$

\end{minipage}\\
\makebox[3em]{$^{57}$}
\begin{minipage}[t]{14cm}
{\it Department of Physics, York University, Ontario, Canada M3J
1P3}~$^{N}$

\end{minipage}\\
\vspace{30em} \pagebreak[4]


\makebox[3ex]{$^{ A}$}
\begin{minipage}[t]{14cm}
 supported by the US Department of Energy\
\end{minipage}\\
\makebox[3ex]{$^{ B}$}
\begin{minipage}[t]{14cm}
 supported by the Italian National Institute for Nuclear Physics (INFN) \
\end{minipage}\\
\makebox[3ex]{$^{ C}$}
\begin{minipage}[t]{14cm}
 supported by the German Federal Ministry for Education and Research (BMBF), under
 contract No. 05 H09PDF\
\end{minipage}\\
\makebox[3ex]{$^{ D}$}
\begin{minipage}[t]{14cm}
 supported by the Science and Technology Facilities Council, UK\
\end{minipage}\\
\makebox[3ex]{$^{ E}$}
\begin{minipage}[t]{14cm}
 supported by an FRGS grant from the Malaysian government\
\end{minipage}\\
\makebox[3ex]{$^{ F}$}
\begin{minipage}[t]{14cm}
 supported by the US National Science Foundation. Any opinion,
 findings and conclusions or recommendations expressed in this material
 are those of the authors and do not necessarily reflect the views of the
 National Science Foundation.\
\end{minipage}\\
\makebox[3ex]{$^{ G}$}
\begin{minipage}[t]{14cm}
 supported by the Polish Ministry of Science and Higher Education as a scientific project No.
 DPN/N188/DESY/2009\
\end{minipage}\\
\makebox[3ex]{$^{ H}$}
\begin{minipage}[t]{14cm}
 supported by the Polish Ministry of Science and Higher Education and its grants
 for Scientific Research\
\end{minipage}\\
\makebox[3ex]{$^{ I}$}
\begin{minipage}[t]{14cm}
 supported by the German Federal Ministry for Education and Research (BMBF), under
 contract No. 05h09GUF, and the SFB 676 of the Deutsche Forschungsgemeinschaft (DFG) \
\end{minipage}\\
\makebox[3ex]{$^{ J}$}
\begin{minipage}[t]{14cm}
 supported by the Japanese Ministry of Education, Culture, Sports, Science and Technology
 (MEXT) and its grants for Scientific Research\
\end{minipage}\\
\makebox[3ex]{$^{ K}$}
\begin{minipage}[t]{14cm}
 supported by the Korean Ministry of Education and Korea Science and Engineering
 Foundation\
\end{minipage}\\
\makebox[3ex]{$^{ L}$}
\begin{minipage}[t]{14cm}
 supported by FNRS and its associated funds (IISN and FRIA) and by an Inter-University
 Attraction Poles Programme subsidised by the Belgian Federal Science Policy Office\
\end{minipage}\\
\makebox[3ex]{$^{ M}$}
\begin{minipage}[t]{14cm}
 supported by the Spanish Ministry of Education and Science through funds provided by
 CICYT\
\end{minipage}\\
\makebox[3ex]{$^{ N}$}
\begin{minipage}[t]{14cm}
 supported by the Natural Sciences and Engineering Research Council of Canada (NSERC) \
\end{minipage}\\
\makebox[3ex]{$^{ O}$}
\begin{minipage}[t]{14cm}
 partially supported by the German Federal Ministry for Education and Research (BMBF)\
\end{minipage}\\
\makebox[3ex]{$^{ P}$}
\begin{minipage}[t]{14cm}
 supported by RF Presidential grant N 4142.2010.2 for Leading Scientific Schools, by the
 Russian Ministry of Education and Science through its grant for Scientific Research on
 High Energy Physics and under contract No.02.740.11.0244 \
\end{minipage}\\
\makebox[3ex]{$^{ Q}$}
\begin{minipage}[t]{14cm}
 supported by the Netherlands Foundation for Research on Matter (FOM)\
\end{minipage}\\
\makebox[3ex]{$^{ R}$}
\begin{minipage}[t]{14cm}
 supported by the Israel Science Foundation\
\end{minipage}\\
\vspace{30em} \pagebreak[4]


\makebox[3ex]{$^{ a}$}
\begin{minipage}[t]{14cm}
now at University of Salerno, Italy\
\end{minipage}\\
\makebox[3ex]{$^{ b}$}
\begin{minipage}[t]{14cm}
now at Queen Mary University of London, United Kingdom\
\end{minipage}\\
\makebox[3ex]{$^{ c}$}
\begin{minipage}[t]{14cm}
also funded by Max Planck Institute for Physics, Munich, Germany\
\end{minipage}\\
\makebox[3ex]{$^{ d}$}
\begin{minipage}[t]{14cm}
also Senior Alexander von Humboldt Research Fellow at Hamburg University,
 Institute of Experimental Physics, Hamburg, Germany\
\end{minipage}\\
\makebox[3ex]{$^{ e}$}
\begin{minipage}[t]{14cm}
also at Cracow University of Technology, Faculty of Physics,
 Mathemathics and Applied Computer Science, Poland\
\end{minipage}\\
\makebox[3ex]{$^{ f}$}
\begin{minipage}[t]{14cm}
supported by the research grant No. 1 P03B 04529 (2005-2008)\
\end{minipage}\\
\makebox[3ex]{$^{ g}$}
\begin{minipage}[t]{14cm}
now at Rockefeller University, New York, NY
 10065, USA\
\end{minipage}\\
\makebox[3ex]{$^{ h}$}
\begin{minipage}[t]{14cm}
now at DESY group FS-CFEL-1\
\end{minipage}\\
\makebox[3ex]{$^{ i}$}
\begin{minipage}[t]{14cm}
now at Institute of High Energy Physics, Beijing, China\
\end{minipage}\\
\makebox[3ex]{$^{ j}$}
\begin{minipage}[t]{14cm}
now at DESY group FEB, Hamburg, Germany\
\end{minipage}\\
\makebox[3ex]{$^{ k}$}
\begin{minipage}[t]{14cm}
also at Moscow State University, Russia\
\end{minipage}\\
\makebox[3ex]{$^{ l}$}
\begin{minipage}[t]{14cm}
now at University of Liverpool, United Kingdom\
\end{minipage}\\
\makebox[3ex]{$^{ m}$}
\begin{minipage}[t]{14cm}
now at CERN, Geneva, Switzerland\
\end{minipage}\\
\makebox[3ex]{$^{ n}$}
\begin{minipage}[t]{14cm}
also affiliated with Universtiy College London, UK\
\end{minipage}\\
\makebox[3ex]{$^{ o}$}
\begin{minipage}[t]{14cm}
now at Goldman Sachs, London, UK\
\end{minipage}\\
\makebox[3ex]{$^{ p}$}
\begin{minipage}[t]{14cm}
also at Institute of Theoretical and Experimental Physics, Moscow, Russia\
\end{minipage}\\
\makebox[3ex]{$^{ q}$}
\begin{minipage}[t]{14cm}
also at FPACS, AGH-UST, Cracow, Poland\
\end{minipage}\\
\makebox[3ex]{$^{ r}$}
\begin{minipage}[t]{14cm}
partially supported by Warsaw University, Poland\
\end{minipage}\\
\makebox[3ex]{$^{ s}$}
\begin{minipage}[t]{14cm}
now at Istituto Nucleare di Fisica Nazionale (INFN), Pisa, Italy\
\end{minipage}\\
\makebox[3ex]{$^{ t}$}
\begin{minipage}[t]{14cm}
now at Haase Energie Technik AG, Neum\"unster, Germany\
\end{minipage}\\
\makebox[3ex]{$^{ u}$}
\begin{minipage}[t]{14cm}
now at Department of Physics, University of Bonn, Germany\
\end{minipage}\\
\makebox[3ex]{$^{ v}$}
\begin{minipage}[t]{14cm}
now at Biodiversit\"at und Klimaforschungszentrum (BiK-F), Frankfurt, Germany\
\end{minipage}\\
\makebox[3ex]{$^{ w}$}
\begin{minipage}[t]{14cm}
also affiliated with DESY, Germany\
\end{minipage}\\
\makebox[3ex]{$^{ x}$}
\begin{minipage}[t]{14cm}
also at University of Tokyo, Japan\
\end{minipage}\\
\makebox[3ex]{$^{ y}$}
\begin{minipage}[t]{14cm}
now at Kobe University, Japan\
\end{minipage}\\
\makebox[3ex]{$^{ z}$}
\begin{minipage}[t]{14cm}
supported by DESY, Germany\
\end{minipage}\\
\makebox[3ex]{$^{\dagger}$}
\begin{minipage}[t]{14cm}
 deceased \
\end{minipage}\\
\makebox[3ex]{$^{aa}$}
\begin{minipage}[t]{14cm}
member of National Technical University of Ukraine, Kyiv Polytechnic Institute,
 Kyiv, Ukraine\
\end{minipage}\\
\makebox[3ex]{$^{ab}$}
\begin{minipage}[t]{14cm}
member of National University of Kyiv - Mohyla Academy, Kyiv, Ukraine\
\end{minipage}\\
\makebox[3ex]{$^{ac}$}
\begin{minipage}[t]{14cm}
partly supported by the Russian Foundation for Basic Research, grant 11-02-91345-DFG\_a\
\end{minipage}\\
\makebox[3ex]{$^{ad}$}
\begin{minipage}[t]{14cm}
Alexander von Humboldt Professor; also at DESY and University of
 Oxford\
\end{minipage}\\
\makebox[3ex]{$^{ae}$}
\begin{minipage}[t]{14cm}
STFC Advanced Fellow\
\end{minipage}\\
\makebox[3ex]{$^{af}$}
\begin{minipage}[t]{14cm}
nee Korcsak-Gorzo\
\end{minipage}\\
\makebox[3ex]{$^{ag}$}
\begin{minipage}[t]{14cm}
This material was based on work supported by the
 National Science Foundation, while working at the Foundation.\
\end{minipage}\\
\makebox[3ex]{$^{ah}$}
\begin{minipage}[t]{14cm}
also at Max Planck Institute for Physics, Munich, Germany, External Scientific Member\
\end{minipage}\\
\makebox[3ex]{$^{ai}$}
\begin{minipage}[t]{14cm}
now at Tokyo Metropolitan University, Japan\
\end{minipage}\\
\makebox[3ex]{$^{aj}$}
\begin{minipage}[t]{14cm}
now at Nihon Institute of Medical Science, Japan\
\end{minipage}\\
\makebox[3ex]{$^{ak}$}
\begin{minipage}[t]{14cm}
now at Osaka University, Osaka, Japan\
\end{minipage}\\
\makebox[3ex]{$^{al}$}
\begin{minipage}[t]{14cm}
also at \L\'{o}d\'{z} University, Poland\
\end{minipage}\\
\makebox[3ex]{$^{am}$}
\begin{minipage}[t]{14cm}
member of \L\'{o}d\'{z} University, Poland\
\end{minipage}\\
\makebox[3ex]{$^{an}$}
\begin{minipage}[t]{14cm}
now at Department of Physics, Stockholm University, Stockholm, Sweden\
\end{minipage}\\
\makebox[3ex]{$^{ao}$}
\begin{minipage}[t]{14cm}
also at Cardinal Stefan Wyszy\'nski University, Warsaw, Poland\
\end{minipage}\\

}


\pagenumbering{arabic} 
\pagestyle{plain}

\section{Introduction}

In exclusive photoproduction
of heavy vector mesons (VMs), $J/\psi$ and
$\Upsilon$, the masses of  the charm and the bottom quarks provide 
a hard scale and the process
can be described by  models based 
on perturbative QCD (pQCD) \cite{pQCD:kmw,pQCD:fks1}.
The interaction
may be viewed at leading order as shown in Fig.~\ref{fig-graphs}:
the photon fluctuates into a  $q\bar{q}$ state of small 
transverse size,
which interacts with partons in the 
proton through a two-gluon colour-singlet 
state, forming a heavy vector meson.
Thus the cross section is proportional to the square of
the gluon density in the proton.
A characteristic feature of heavy  VM photoproduction  
is the rapid  rise of the  cross section with
the photon--proton centre-of-mass energy, $W$.  
This  can be explained through
the increasing  gluon density with  decreasing fractional momentum, 
$x\propto 1/W^2$ (where the $x$ region accessible
in heavy-quark production at HERA is $10^{-4}<x<10^{-2}$).
Numerous studies have shown that 
the dependence of the cross section  
on  $W$ can be  parameterised as $\sigma
\propto W^\delta$ \cite{GWolf,proc:Bruni-Janssen-Marage}.
Measurements for the $J/\psi$ meson~\cite{epj:c24:345,epj:c46:585} 
yielded  $\delta\approx 0.7$.  
A higher value of
$\delta \approx 1.7$ has been  predicted for exclusive photoproduction 
of $\Upsilon(1S)$ mesons
in leading-order pQCD~\cite{Frankfurt:1998yf}, 
consistent with the  recent ZEUS measurement:  
$\delta = 1.2 \pm 0.8$~\cite{Chekanov:2009zz}.

Studies of the exclusive photoproduction of light and heavy vector 
mesons \cite{GWolf} have shown that  the $t$ dependence
of the differential cross section may be approximated 
in the region of small $t$
($|t| < 1$~\gev$^2$) with a single exponential: 
$d\sigma / d|t| \propto \exp{(-b|t|)}$, where $t$
is the four-momentum-transfer squared at the proton vertex.  
The slope parameter, $b$, measured at ZEUS
for exclusive $J/\psi$ production~\cite{epj:c24:345} at $W_0 = 90$~\gev~is   
$b = 4.15\pm 0.05$ (stat.) $^{+0.30}_{-0.18}$ (syst.)~\gev$^{-2}$ 
and exhibits a logarithmic  variation:
$b(W) = b_0 + 2\alpha'\ln(W/W_0)^2$, where $\alpha' \approx 0.1$~\gev$^{-2}$.
In an optical model approach for exclusive production
of VMs, the slope parameter $b$ is related to the 
radii of the  proton, $R_{p}$, and  the vector meson, $R_{VM}$,
according to the approximate formula:
$b \approx (R^2_p + R^2_{VM})/4$. 
The value of $b$ measured for $J/\psi$ production is approximately
equal to that expected from the size of the proton
($b \approx 4~{\rm \gev}^{-2}$), in agreement with calculations
based on pQCD \cite{pQCD:fks2}.
This suggests that the size of the 
$J/\psi$ is small compared to that of the proton.
A similar picture is expected in the case of exclusive 
$\Upsilon(1S)$ production \cite{PanJi-Huan,Cox:DiffY}.



The present paper reports on the first measurement of
$b$ in exclusive $\Upsilon(1S)$ photoproduction,
observed in the $\mu^+\mu^-$ decay channel
in the kinematic range $60<W<220$~\gev,
and complements the previous 
results~\cite{Chekanov:2009zz,pl:b437:432,pl:b483:23} 
on $\Upsilon(1S)$ photoproduction. 
The data correspond to an 
integrated luminosity of $468$~pb$^{-1}$,
collected in the period 1996--2007.


\section{Experimental set-up}
In 1998--2007 (1996--1997), HERA provided electron\footnote{Electrons
and positrons are both referred to as electrons in this article.} beams
of energy $E_e = 27.5$~\gev~and proton beams of energy $E_p = 920\ (820)$~\gev,
resulting in a centre-of-mass energy of $\sqrt s=318\ (300)$~\gev.

\Zdetdesc


In the kinematic range of the analysis, charged particles were tracked
in the central tracking detector (CTD)~\citeCTD and, for the data
taken after 2001, also in the microvertex detector
(MVD)~\citeMVD. These components operated in a magnetic field of
$1.43\Tesla$ provided by a thin superconducting solenoid. The CTD
consisted of 72~cylindrical drift chamber layers, organised in nine
superlayers covering the polar-angle\footnote{The ZEUS coordinate
system is a right-handed Cartesian system, with the $Z$ axis pointing
in the proton beam direction, referred to as the ``forward
direction'', and the $X$ axis pointing left towards the centre of
HERA.  
The coordinate origin was located at the nominal interaction point
for data collected before 2001. After 2001 it was redefined
as the centre of the CTD.
The polar angle, $\theta$, is measured with respect to the proton beam
direction.\xspace}
region
\mbox{$15^\circ<\theta<164^\circ$}. 
The MVD provided polar angle coverage from $7^\circ$
to $150^\circ$.
The transverse-momentum resolution for full-length tracks was
$\sigma(p_T)/p_T = 0.0058\,p_T \oplus 0.0065 \oplus 0.0014/p_T$,
with $p_T$ in~\gev, for data taken before 2001 and
$\sigma(p_T)/p_T = 0.0029\,p_T \oplus 0.0081 \oplus 0.0012/p_T$,
for data taken after 2001.

\Zcaldesc

The muon system consisted of barrel, rear (B/RMUON)~\cite{brmuon} and
forward (FMUON)~\cite{zeus:1993:bluebook} tracking detectors. The
B/RMUON consisted of limited-streamer (LS) tube chambers placed behind
the BCAL (RCAL), both inside and outside the magnetised iron yoke
surrounding the CAL. The barrel and rear muon chambers covered polar
angles from 34$^{\circ}$ to 135$^{\circ}$ and from 135$^{\circ}$ to
171$^{\circ}$, respectively.  The FMUON consisted of six planes of LS
tubes and four planes of drift chambers covering the angular region
from 5$^{\circ}$ to 32$^{\circ}$.  The muon system exploited the
magnetic field of the iron yoke and, in the forward direction, of two
iron toroids magnetised to 1.6~T to provide an independent measurement
of the muon momentum.

\Zbacdesc



The luminosity was measured using the Bethe-Heitler reaction
$ep \rightarrow e\gamma \,p$ with the luminosity detector which
consisted of independent lead--scintillator
calorimeter~\cite{acpp:b32:2025} and magnetic
spectrometer~\cite{nim:a565:572} systems.


\section{Kinematics}

\label{section-kinematics}

\begin{figure}[p]
\vfill
\begin{center}
\includegraphics[width=0.72\textwidth,angle=0]{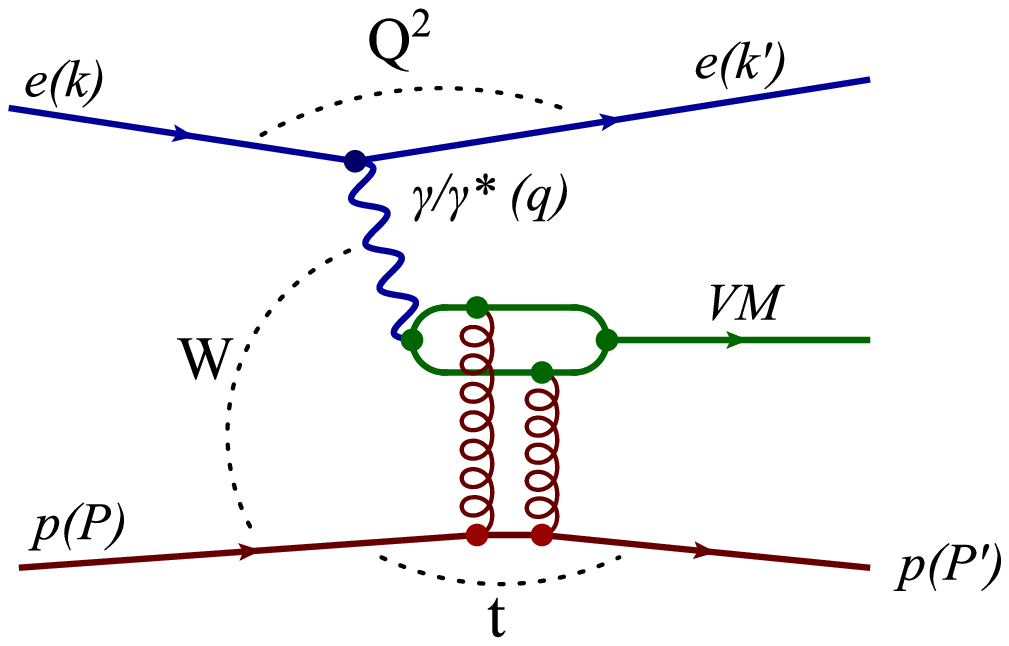}\\[2.4cm]
\includegraphics[width=0.72\textwidth,angle=0]{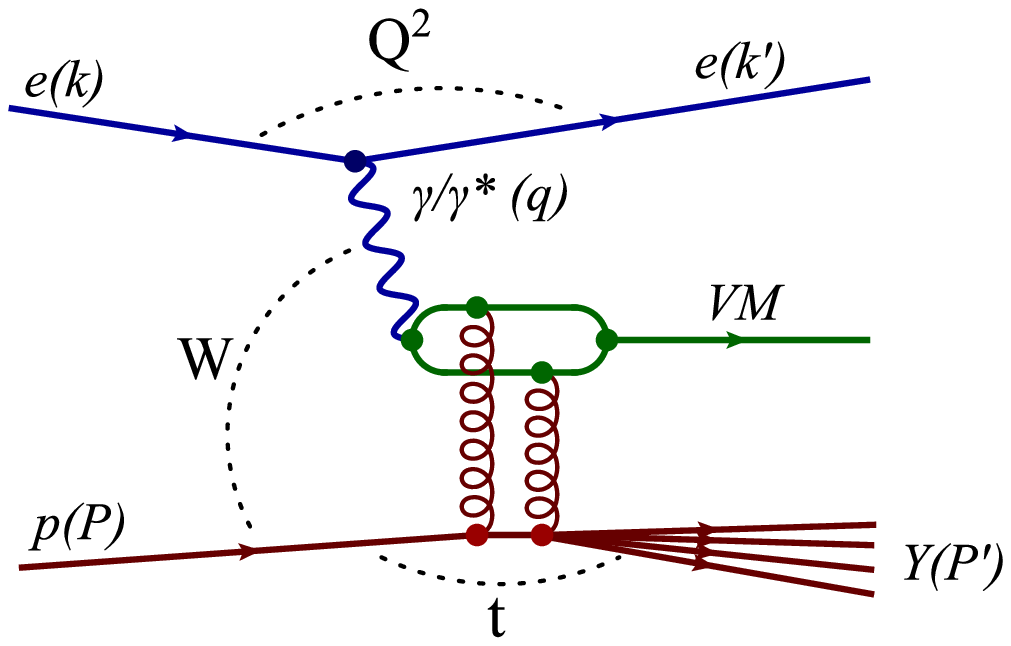}\\[1.4cm]
\end{center}
\caption{Diagrams for (top) exclusive and (bottom) proton-dissociative
vector-meson photoproduction in $ep$ interactions.
The variables describing the kinematics of both processes are introduced in
Section \ref{section-kinematics}.
}
\label{fig-graphs}
\vfill
\end{figure}

The four-momenta of the incoming and outgoing electron and proton 
are denoted by $k,k',P$ and $P'$,
respectively.
The exclusive reaction under study
\begin{equation}
ep \rightarrow e\Upsilon p \rightarrow e \mu^+\mu^- p
\end{equation}
is described by the following variables (Fig.~\ref{fig-graphs}, top):
\begin{itemize}
\item
  $s=(k+P)^2$, the centre-of-mass-energy squared of the
electron--proton system; 
\item $Q^2 = -q^2 = -(k-k')^2$, 
the negative four-momentum squared of the exchanged photon;
\item $y=(q\cdot P)/(k\cdot P)$, the fraction of 
the  electron energy transferred to
the hadronic final state in the rest frame of the initial-state proton;
\item $W^2=(q+P)^2=-Q^2+2y(k\cdot P)+m_p^2$, 
the centre-of-mass-energy squared of the
photon--proton system, where $m_p$ is the proton mass;
\item
  $M_{\mu^+ \mu^-}$, the invariant mass of the $\mu^+ \mu^-$ pair;
\item
  $t=(P-P')^2$, the squared four-momentum transfer at the proton vertex,
  determined from the approximate formula: 
$t \approx -(p_{x}^{+} + p_{x}^{-})^2 - (p_{y}^{+} + p_{y}^{-})^2$,
where $p_{x,y}^{\pm}$ are the components of the transverse momentum 
of the decay muons.
\end{itemize}
The reaction $ep \rightarrow e\Upsilon Y$, (Fig.~\ref{fig-graphs}, bottom),
where $Y$ denotes a hadronic state originating from proton dissociation,
constitutes an important background.
These events mimic exclusive $\Upsilon$ production when
the hadrons from proton dissociation remain undetected.

Events used in the analysis were restricted to $Q^2$ values from
the kinematic minimum, $Q^2_{\rm min}=m_e^2y^2/(1-y) \approx
10^{-9}$~\gev$^2$, where $m_e$ is the electron mass, to a
value at which the scattered electron starts to be observed in the
CAL, $Q^2_{\rm max}\approx$ 1~\gev$^2$,
with an estimated median $Q^2$ value of $10^{-3}$ \gev$^2$. 
The photon--proton centre-of-mass energy can then be expressed as:
\begin{equation}
W^2 \approx 4E_p E_e y \approx 2E_p(E-p_Z),
\label{W-approx}
\end{equation}
where $E$ is the energy and $p_Z$ is
the longitudinal momentum of the $\mu^+\mu^-$ pair.

The approximate formula for $t$ introduces dispersion $3$ times
smaller then that in the experimental resolution of this variable
after all event selections;
approximation (\ref{W-approx}) has a negligible effect in the case of $W$. 
\section{Event selection}

Exclusive $\mu^+ \mu^-$ events in photoproduction 
were selected online by requiring at least one CTD track
associated with a F/B/RMUON
deposit or with a signal in the BAC consistent with a muon.
Owing to the inclusion of muon triggers based on signals
in the BAC~\cite{bac:1,thesis:plucinski},
the rate of recorded dimuon events increased by $17\%$ for 
a third of the data as compared to the previous $\Upsilon(1S)$
analysis~\cite{Chekanov:2009zz}.
Offline, events were selected as follows:

\begin{itemize}
\item two oppositely
charged tracks forming a vertex and no other tracks 
present in the central tracking system;
\item position of the vertex consistent with an $ep$ interaction;
\item both tracks were required to have
hits in at least 5 CTD superlayers, to ensure a good
momentum resolution;
\item
  transverse momentum of each track  $p_T > 1.5$~\gev;
\item $|\eta^+ - \eta^-|\leq$~1.5, where $\eta^{\pm}$ 
is the pseudorapidity\footnote{
Pseudorapidity is defined as $\eta = -\ln{(\tan\frac{\theta}{2})}$.} of
a given track, to suppress Bethe-Heitler background (Section \ref{MC-simul});
\item
at least one track identified as a muon in the F/B/RMUON or
BAC, whenever available in a given 
event~\citeRubinskyMalka; if not
explicitly identified as a muon, the second track had to be
associated with a minimum-ionising energy deposit in the CAL;
\item
$|\pi-\alpha|>0.1$, where $\alpha$ is the angle between the momentum
vectors of $\mu^+$ and $\mu^-$,
to reject cosmic-ray events;
\item invariant mass $M_{\mu^+\mu^-}$ in the range between 5 and 15 \gev;
\item the energy of each CAL cluster not associated to any of
the final-state muons was required to be less than
$0.5$~\gev, in order to be above the noise level of the CAL. 
It implicitly selected exclusive events with
an effective cut $Q^2<1$~\gev$^2$;
\item the sum of the energy in the FCAL modules surrounding the beam hole 
had to be smaller than $1$~\gev~\citeRubinskyMalka
to suppress the contamination from
proton-dissociative events, $ep \rightarrow e\Upsilon Y$. 
According to a Monte Carlo study,
this corresponds to an effective cut on
the mass $M_Y$ of the dissociated system originating from the proton, 
$M_Y \lesssim$~4~\gev;
\item photon--proton centre-of-mass energy $60<W<220$~\gev~and 
four-momentum-transfer squared $|t|<5$~\gev$^2$.
\end{itemize}
The total number of selected $\mu^+\mu^-$ pairs was 2769.
The contamination of this sample with cosmic ray muons is less then
1\%.
\section{Monte Carlo simulation}

\label{MC-simul}

The detector and trigger acceptance and the effects due to
detector response were determined using samples of Monte Carlo (MC)
events.
Exclusive and proton-dissociative vector-meson  
production were simulated with the DIFFVM 2.0
generator~\cite{proc:mc:1998:396}.
For proton-dissociative events, the simulation was supplemented 
by the JETSET 7.3 MC 
package~\cite{manual:cern-th-7112/93}.
For exclusive vector-meson
production, $s$-channel helicity conservation (SCHC) was assumed. 
An exponential dependence, $e^{-b|t|}$, was assumed for the differential
cross section in $t$ with a slope parameter $b = 4.5$
\gev$^{-2}$, consistent with the value obtained for exclusive $J/\psi$
electroproduction~\cite{epj:c24:345,epj:c46:585}. The $W$ dependence
of the $\gamma p \rightarrow \Upsilon p$ cross section was
parameterised as $\propto W^\delta$, 
with $\delta$ = 1.2~\cite{Chekanov:2009zz}.
Electromagnetic radiative corrections associated with the decay muons 
are of the order of 1\,\%~\cite{spiridonov-2005}
and were not included in the simulation. 

The non-resonant background, consisting of the exclusive 
and proton-dissociative 
Bethe-Heitler (BH) dimuon events, was simulated using the GRAPE v1.1k
MC program~\cite{cpc:136:126}.  
After event selection, the contribution of the proton-dissociative events 
was 25\% of the Bethe-Heitler MC sample.

All MC events were generated in the full kinematic range and
processed through the
simulation of the ZEUS detector based on the GEANT
program\footnote{Version 3.13 for the 1996--2000 and
3.21 for the 2003--2007 periods, respectively.}~\cite{tech:cern-dd-ee-84-1}
and were analysed with the same reconstruction and
offline procedures as the data.  In addition,
corrections~\citeRubinskyMalka of the muon-detector
efficiencies determined from a data set consisting of 
$J/\psi$ and Bethe-Heitler exclusive production events were applied.

\section{Determination 
of the \boldmath{$b$} slope}

\begin{figure}[p]
\vfill
\begin{center}
\includegraphics[width=0.98\textwidth,angle=0]{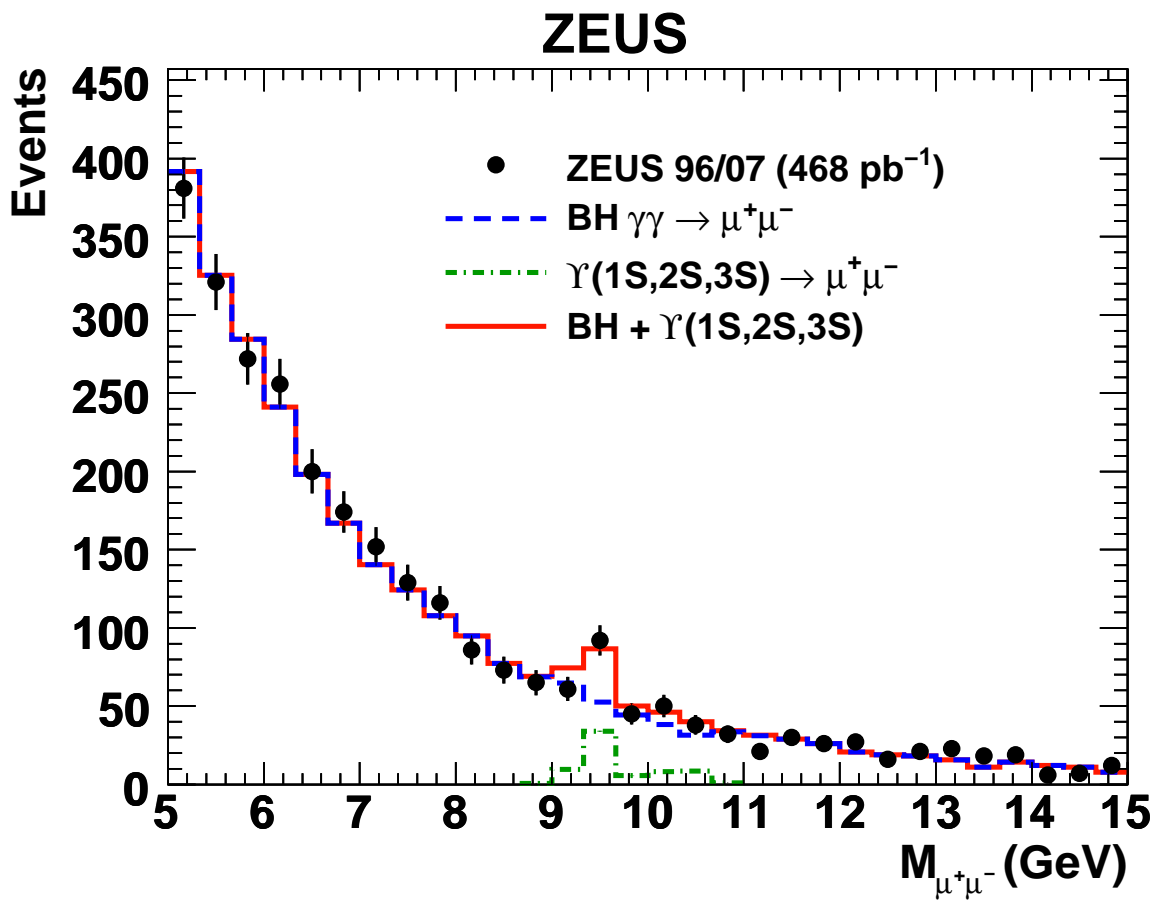}
\end{center}
\caption{
Invariant mass distribution of $\mu^+\mu^-$ pairs. 
The  dashed line shows the  simulated Bethe-Heitler (BH)
(exclusive and proton dissociative) distribution,   
normalised to the data points
in the range [5.0--15.0]~\gev~excluding the [9.0--11.0]~\gev~mass window. 
Simulated contributions of 
the $\Upsilon(1S)$, $\Upsilon(2S)$ and $\Upsilon(3S)$ resonances are  
shown as a histogram on the mass  axis (dashed-dotted line).  
The solid line shows the sum of all contributions.}
\label{fig-m_inv}
\vfill
\end{figure}

The invariant-mass distribution of $\mu^+\mu^-$  pairs after
applying the selection criteria
is shown in Fig.~\ref{fig-m_inv}.
The simulated contributions from the Bethe-Heitler 
(exclusive and proton dissociative) process and from the $\Upsilon(1S)$,
$\Upsilon(2S)$ and $\Upsilon(3S)$ resonances are also 
presented\footnote{The ratio of the number
of events $N_{\Upsilon(1S)}:N_{\Upsilon(2S)}:N_{\Upsilon(3S)}$ 
was fixed in the MC to $0.73:0.19:0.08$
according to a CDF measurement \cite{CDFups02} of the production 
of $\Upsilon$ resonances.}.
As in the previous paper~\cite{Chekanov:2009zz}, the BH distributions
were normalised to the data 
in the range [5.0--15.0]~\gev~excluding the [9.0--11.0]~\gev~mass window
where contributions from the $\Upsilon$ resonances are expected.
For the determination of the slope parameter 
for exclusive $\Upsilon(1S)$ production, 
only events in the mass window [9.33--9.66]~\gev~were considered.
The width of the mass window was chosen
in order to avoid excessive smearing of the $t$ variable
and to retain a good signal-to-background ratio.
According to MC studies, 71\%  
of all reconstructed $\Upsilon(1S)$ events  are expected
in this window;
the relative contaminations of $\Upsilon(2S)$ and
$\Upsilon(3S)$ states with respect to $\Upsilon(1S)$ 
are 1.3\% and 0.1\%, respectively.
The contribution from the $\Upsilon(2S)$ and $\Upsilon(3S)$ 
states was neglected for the extraction of the slope parameter $b$.
After scanning no cosmic ray muon candidates were found
in the signal mass window.

\begin{figure}[p]
\vfill
\begin{center}
\includegraphics[width=1.00\textwidth,angle=0]{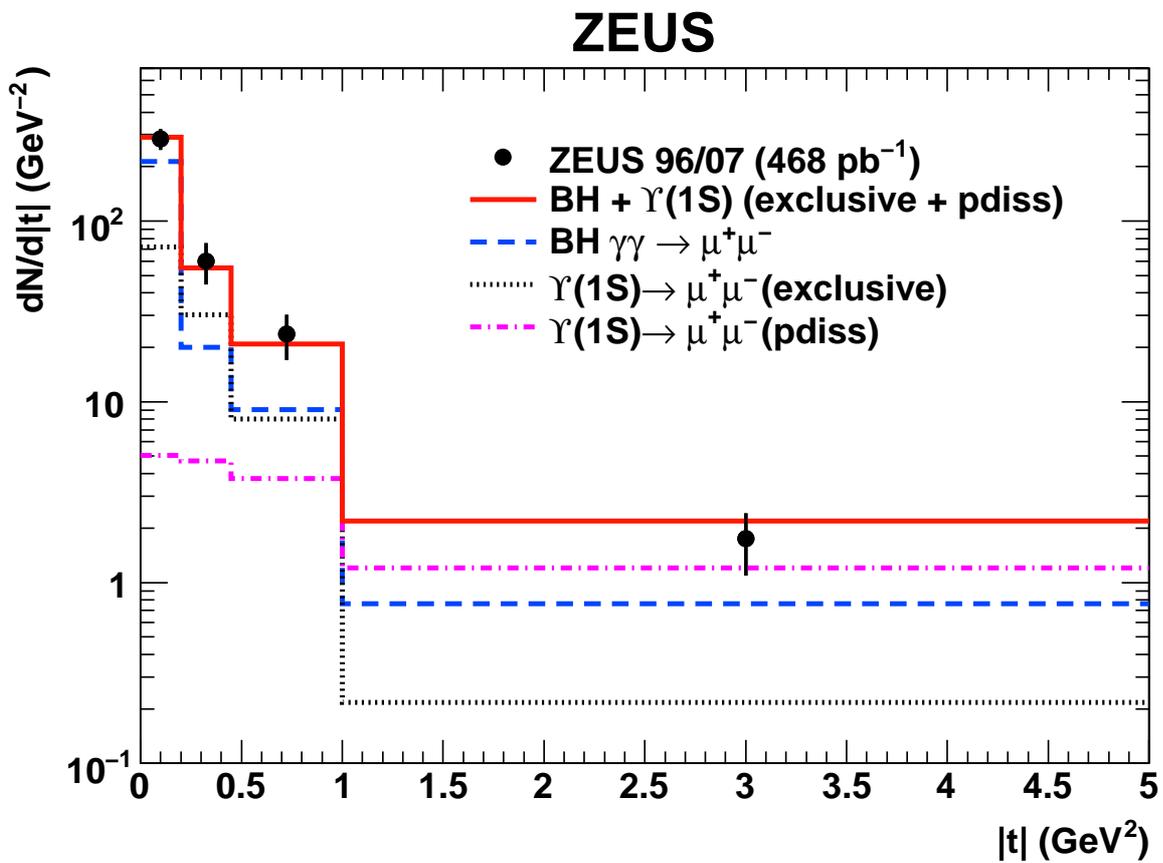}
\end{center}
\caption{
Measured $|t|$ distribution (full dots) with  error bars 
denoting statistical uncertainties. 
Fitted distributions for simulated events are shown for the Bethe-Heitler
(dashed line), exclusive $\Upsilon(1S)$ (dotted line) and 
proton dissociative $\Upsilon(1S)$ (dashed-dotted line) processes.
The solid line shows the sum of all contributions.}
\label{fig-dN_dt}
\vfill
\end{figure}

The value of the slope parameter for exclusive $\Upsilon(1S)$
production was determined as follows:
the sum of simulated distributions of all
contributing processes was fitted to the observed
event yields in the signal mass window [9.33--9.66]~\gev~in the 
four $t$ bins shown in Fig.~\ref{fig-dN_dt}. A binned Poissonian
log-likelihood function, $\ln{(L)}$, was used.
The expected number of Bethe-Heitler background events was fixed
to the value obtained from the $\mu^+\mu^-$ spectrum
outside the signal region as described earlier.
Due to insufficient statistics
it was not  possible  to evaluate the contribution of 
proton-dissociative $\Upsilon(1S)$ events in the final sample 
with the present data.
However, the fraction  of such  events, $f_{\rm pdiss}$, 
is expected to be similar in all diffractive vector-meson production
processes~\cite{epj:c14:213}. 
Therefore, the value $f_{\rm pdiss}=0.25 \pm 0.05$, determined for
diffractive $J/\psi$ production ~\cite{epj:c24:345}, was used.
The values of the slope parameter for the exclusive and proton
dissociative $\Upsilon(1S)$ production processes differ \cite{Aaron:2009xp};
in the MC the value for the latter was taken to be 
$b_{\rm pdiss} = 0.65 \pm 0.1$~\gev$^{-2}$~\cite{epj:c24:345}.

The fit was performed with two free parameters: the slope $b$ 
and the number of expected 
$\Upsilon(1S)$ events in the 
signal mass window.
During the parameter scan, the contribution of the exclusive $\Upsilon(1S)$ 
production to the $t$ distribution was reweighted at generator level
to the function $b\cdot\exp{(-b|t|)}$.
The small statistical uncertainties of the MC sample 
were neglected in the fit.
The fit yielded:
$b=4.3^{+2.0}_{-1.3}$ (stat.)~\gev$^{-2}$
and $41 \pm 10$ $\Upsilon(1S)$ events (44\% of the events in this mass window).
The fit provides a good description of the data;
the equivalent $\chi^2$ is $0.61$ for $2$ degrees of freedom.

\section{Systematic uncertainties}

The following sources of systematic uncertainty were considered,
where the numbers in parenthesis correspond to the
uncertainties on $b$ in \gev$^{-2}$:
\begin{itemize}
\item $f_{\rm pdiss}$ was varied between 0.2 and 0.3,
as determined from
$J/\psi$ production~\cite{epj:c24:345} ($^{+0.30}_{-0.25}$);
\item $b_{\rm pdiss}$  was varied 
by $^{+0.7}_{-0.1}$~\gev$^{-2}$.
In addition to the uncertainty from $J/\psi$ production
quoted earlier, the upper variation was extended 
to the value $b_{\rm pdiss} = 1.35$~\gev$^{-2}$ obtained
for this parameter when it was also fitted ($^{-0.4}_{+0.1}$);
\item  the contribution of BH events 
in the mass window 
[9.33--9.66]~\gev~was varied between $55.3\%$ and $56.7\%$,
according to the statistical uncertainty of its normalisation
($^{+0.15}_{- 0.10}$);
\item 
the fraction of proton-dissociative to all BH events
was varied in the range 0.22 to 0.28 ($\pm 0.30$).
\end{itemize}
Variation of the parameter $\delta$ between $0.7$ and $1.7$ and
variations of the offline selection cuts lead to a negligible contribution to
the $b$ uncertainty.
The total systematic uncertainty was determined by adding the
individual contributions in quadrature.

\section{Result and discussion}

\begin{figure}[p]
\vfill
\begin{center}
\includegraphics[width=1.00\textwidth]{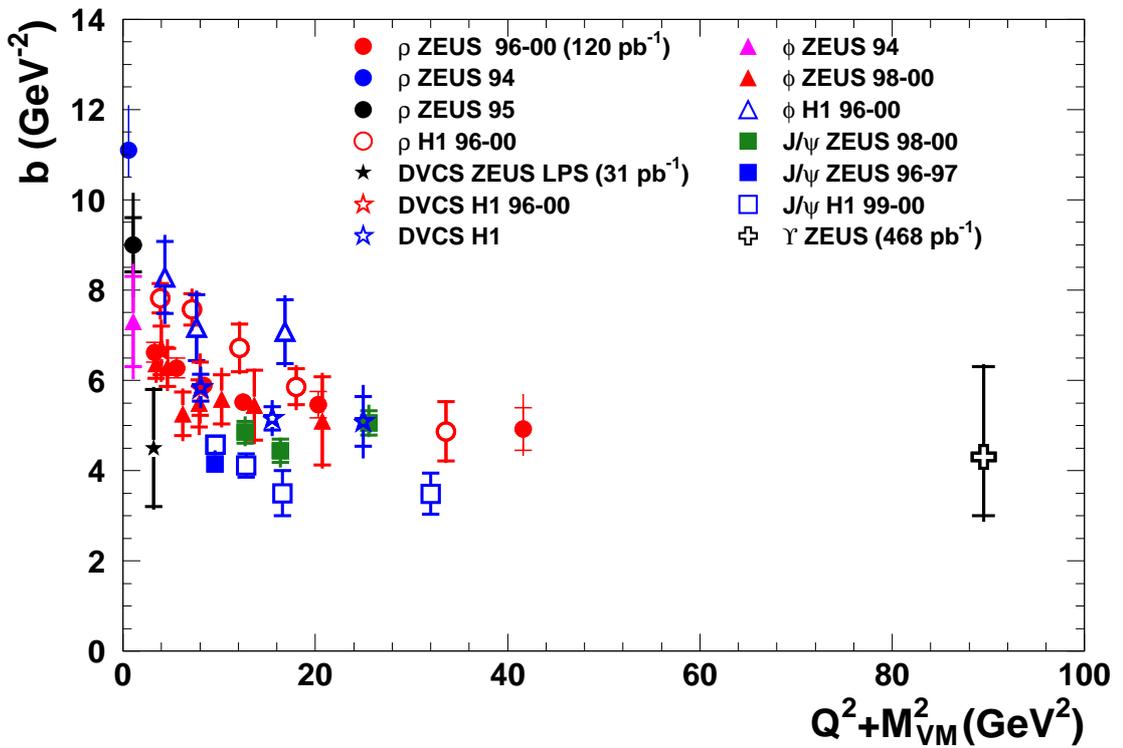}
\end{center}
\caption{
Comparison of the HERA measurements of the slope parameter $b$  
as a function of the scale $Q^2 + M_{VM}^2$
for exclusive $\Upsilon(1S)$ production (the rightmost data point), 
for other exclusive vector-meson 
production~\protect\cite{pmcphysics:a1:6,epj:c2:247,epj:c6:603,Aaron:2009xp,
pl:b377:259,np:b718:3,np:b695:3,epj:c24:345,epj:c46:585}
and for deeply virtual Compton scattering 
(DVCS)~\protect\cite{jhep:05:108,Aktas:2005ty,Aaron:2007cz}.}
\label{fig-b_shape_lin}
\vfill
\end{figure}

The slope parameter $b$ for the exclusive production 
of $\Upsilon(1S)$ mesons was measured to be
$b=4.3^{+2.0}_{-1.3}$ (stat.)$\, ^{+0.5}_{-0.6}$ (syst.)~\gev$^{-2}$.
A comparison of all HERA measurements of the slope parameter $b$ for
exclusive light and heavy vector meson production and
for deeply virtual Compton scattering (DVCS)
is shown in Fig.~\ref{fig-b_shape_lin}.
This analysis doubles the range  covered by previous
measurements in terms of $Q^2 + M_{VM}^2$,
where $M_{VM}$ denotes the mass of a vector meson.
The measured value is in agreement with an asymptotic behaviour of this
dependence, reflecting the proton radius.
This was already suggested by earlier measurements and is
consistent with predictions based on pQCD models
($b=3.68$~\gev$^{-2}$) \cite{Cox:DiffY}.

\section{Conclusions}

The exclusive photoproduction reaction 
$\gamma \,p \rightarrow \Upsilon(1S)\,p$ was studied
with the ZEUS detector in $ep$ collisions
at HERA using an integrated luminosity of 468~pb$^{-1}$ 
collected in the period 1996--2007.
The analysis covered  the kinematic range 
$60<W<220$~\gev~and $Q^2<1$ \gev$^2$.
The measurement of the 
exponential slope of the $t$ dependence yielded
$b=4.3^{+2.0}_{-1.3}$ (stat.)$\, ^{+0.5}_{-0.6}$ (syst.)~\gev$^{-2}$. 
This is the first determination of the $b$ parameter for $\Upsilon(1S)$ 
production.
The result is in agreement with expectations of an asymptotic 
behaviour of the slope parameter as a function 
of the effective scale present in the process, $Q^2 + M_{VM}^2$. 
This measurement extends the value of the scale to 
$\approx 90$~\gev$^2$, the highest achieved
to date in the measurement of the $t$-slope parameter
for a vector meson. 

\section*{Acknowledgments}
We appreciate the contributions to the construction and maintenance of
the ZEUS detector of many people who are not listed as authors. The
HERA machine group and the DESY computing staff are especially
acknowledged for their success in providing excellent operation of the
collider and the data-analysis environment. We thank the DESY
directorate for their strong support and encouragement.

\clearpage
{\raggedright
\providecommand{\etal}{et al.\xspace}
\providecommand{\coll}{Collaboration}
\catcode`\@=11
\def\@bibitem#1{%
\ifmc@bstsupport
  \mc@iftail{#1}%
    {;\newline\ignorespaces}%
    {\ifmc@first\else.\fi\orig@bibitem{#1}}
  \mc@firstfalse
\else
  \mc@iftail{#1}%
    {\ignorespaces}%
    {\orig@bibitem{#1}}%
\fi}%
\catcode`\@=12

}



%
%
\end{document}